\documentclass[a4paper,UKenglish,cleveref, autoref]{lipics-v2019}

\usepackage{amsmath}

\hideLIPIcs

\newcommand{\FPT}{\mbox{\small\rm FPT}}

\newtheorem{obs}{Observation} 
\newcommand{\smallp}{\mbox{\small\rm P}}

\newcommand{\Det}{\mbox{\small\rm Det}}

\newcommand{\w}{\mbox{\rm W}}

\usepackage{thm-restate}
\usepackage{hyperref}

\newcommand{\ex}{\mbox{\small\rm X}} 
\newcommand{\ey}{\mbox{\small\rm Y}} 
\newcommand{\ez}{\mbox{\small\rm Z}} 
\newcommand{\daF}{\downarrow \hspace{-0.08 cm}F}

\newcommand{\ABP}{\mbox{\small\rm ABP}}

\newcommand{\F}{\mathbb{F}}

\renewcommand{\angle}[1]{{\langle} #1 {\rangle}}

\newcommand{\sgn}{\mbox{\small\rm sgn}}

\DeclareMathOperator{\poly}{\mbox{\small\rm poly}}

 \DeclareMathOperator{\rper}{\mbox{\rm rPer}} 
\DeclareMathOperator{\per}{\mbox{\rm Per}}
\DeclareMathOperator{\rdet}{\mbox{\rm rDet}} 

\title{On Explicit Branching Programs for the Rectangular Determinant
  and Permanent Polynomials} 

\titlerunning{On Explicit Branching Programs}

\author{V. Arvind}{Institute of Mathematical Sciences (HBNI), Chennai, India}{email: arvind@imsc.res.in}{}{} 
\author{Abhranil Chatterjee}{Institute of Mathematical Sciences (HBNI), Chennai, India}{email: abhranilc@imsc.res.in}{}{} 
\author {Rajit Datta}{Chennai Mathematical Institute, Chennai, India}{email: rajit@cmi.ac.in}{}{}
\author {Partha Mukhopadhyay}{Chennai Mathematical Institute, Chennai, India}{email: partham@cmi.ac.in}{}{}

\authorrunning{Arvind et. al.} 

\Copyright{John Q. Public and Joan R. Public}

\ccsdesc[100]{Theory of Computation~Computational Complexity}
\ccsdesc[100]{Circuit Complexity}

\keywords{Determinant, Permanent, Parameterized Complexity, Branching Programs}

\category{}

\relatedversion{}

\supplement{}

\nolinenumbers 

\EventEditors{John Q. Open and Joan R. Access}
\EventNoEds{2}
\EventLongTitle{42nd Conference on Very Important Topics (CVIT 2016)}
\EventShortTitle{CVIT 2016}
\EventAcronym{CVIT}
\EventYear{2016}
\EventDate{December 24--27, 2016}
\EventLocation{Little Whinging, United Kingdom}
\EventLogo{}
\SeriesVolume{42}
\ArticleNo{23}

\begin{document}

\maketitle

\begin{abstract}
  We study the arithmetic circuit complexity of some well-known family
  of polynomials through the lens of parameterized complexity. Our
  main focus is on the construction of explicit algebraic branching
  programs (ABP) for determinant and permanent polynomials of the
  \emph{rectangular} symbolic matrix in both commutative and
  noncommutative settings. The main results are:

\begin{itemize}

\item We show an explicit $O^{*}({n\choose {\downarrow k/2}})$-size
  ABP construction for noncommutative permanent polynomial of $k\times
  n$ symbolic matrix.  We obtain this via an explicit ABP construction
  of size $O^{*}({n\choose {\downarrow k/2}})$ for $S_{n,k}^*$,
  noncommutative symmetrized version of the elementary symmetric
  polynomial $S_{n,k}$.


\item We obtain an explicit $O^{*}(2^k)$-size ABP construction for the
  commutative rectangular determinant polynomial of the $k\times n$
  symbolic matrix.

\item In contrast, we show that evaluating the rectangular noncommutative
  determinant over rational matrices is $\#\w[1]$-hard.

\end{itemize}

\end{abstract}

\section{Introduction}\label{intro}

The complexity of arithmetic computations is usually studied in the
model of arithmetic circuits and its various restrictions. An
\emph{arithmetic circuit} is a directed acyclic graph with each
indegree-$0$ node (called an input gate) labeled by either a variable
in $\{x_1,x_2,\ldots,x_n\}$ or a scalar from the field $\F$, and all
other nodes (called gates) labeled as either $+$ or $\times$ gate. At
a special node (designated the output gate), the circuit computes a
multivariate polynomial in $\F[x_1,x_2,\ldots,x_n]$. Usually we use
the notation $\F[\ex]$ to denote the polynomial ring
$\F[x_1,x_2,\ldots,x_n]$.

Arithmetic computations are also considered in the noncommutative
setting.  The free noncommutative ring $\F\angle{y_1,y_2, \ldots,
  y_n}$ is usually denoted by $\F\angle{\ey}$\footnote{Throughout the paper, we use $\ex$ to denote the set of commuting variables and $\ey$ to denote the set of noncommuting variables. }. In the ring
$\F\angle{\ey}$, monomials are words in $\ey^*$ and polynomials in
$\F\angle{\ey}$ are $\F$-linear combinations of words. We define
noncommutative arithmetic circuits essentially as their commutative
counterparts.  The only difference is that at each product gate in a
noncommutative circuit there is a prescribed left to right ordering of
its inputs.

A more restricted model than arithmetic circuits are algebraic
branching programs. An \emph{algebraic branching program} (ABP) is a
directed acyclic graph with one in-degree-$0$ vertex called
\emph{source}, and one out-degree-$0$ vertex called \emph{sink}. The
vertex set of the graph is partitioned into layers $0,1,\ldots,\ell$,
with directed edges only only between adjacent layers ($i$ to
$i+1$). The source and the sink are at layers zero and $\ell$
respectively. Each edge is labeled by a linear form over variables
$x_1,x_2,\ldots,x_n$. The polynomial computed by the ABP is the sum
over all source-to-sink directed paths of the product of linear forms
that label the edges of the path. An ABP is \emph{homogeneous} if all
edge labels are homogeneous linear forms. ABPs can be defined in both
commutative and noncommutative settings.

The main purpose of the current paper is to present new arithmetic
complexity upper bound results, in the form of ``optimal'' algebraic
branching programs, for some important polynomials in both the
commutative and noncommutative domains. These results are motivated by
our recent work on an algebraic approach to designing efficient
parameterized algorithms for various combinatorial problems~\cite{ACDM18}.

We now proceed to define the polynomials and explain the results
obtained.

\subsubsection*{\textbf The Elementary Symmetric Polynomial}

We first recall the definition of $k^{th}$ elementary
symmetric polynomial $S_{n,k}\in\F[\ex]$, over the $n$ variables
$\ex=\{x_1,x_2,\ldots,x_n\}$,
\[
 S_{n,k}(\ex)= \sum_{S\subseteq [n] : |S|=k}\prod_{i\in S} x_i.
\]
It is well-known that $S_{n,k}(\ex)$ can be computed by an algebraic
branching program of size $O(nk)$.  In this paper, we consider the
noncommutative symmetrized version $S^*_{n,k}$, in the ring
$\F\angle{\ey}$, defined as:
\[
S^{*}_{n,k}(\ey) = \sum_{T\subseteq [n] : |T|=k} \sum_{\sigma\in S_k}
\prod_{i\in T} y_{\sigma(i)}.
\]       
The complexity of the polynomial $S_{n,k}^{*}$ is first considered by
Nisan in his seminal work in noncommutative computation~\cite{Ni91}. Nisan shows that any ABP for $S_{n,k}^{*}$ is of size
$\Omega(\binom{n}{\downarrow k/2})$~~\footnote{We use
  $\binom{n}{\downarrow r}$ to denote $\sum_{i=0}^r \binom{n}{i}$.}.
Furthermore, Nisan also shows the \emph{existence} of ABP of size
$O(\binom{n}{\downarrow k/2})$ for $S^*_{n,k}$. However, it is not
clear how to construct such an ABP in time $O(\binom{n}{\downarrow
  k/2})$. Note that an ABP of size $O^{*}(n^{k})$ for $S_{n,k}^{*}$
can be directly constructed in $O^{*}(n^k)$ time by opening up the
expression completely \footnote{In this paper we use the notation
  $O^{*}(\cdot)$ freely to suppress the terms asymptotically smaller than
  the main term.}. The main upper bound question is whether we can
achieve any constant factor saving of the parameter $k$ in terms of
size and run time of the construction. In this paper, we give such an
explicit construction.
Note that Nisan's result also rules out any $\FPT(k)$-size ABP for
$S_{n,k}^{*}$. That also justifies the problem from an \emph{exact
  computation} point of view.

\subsubsection*{Rectangular
  Permanent and Rectangular Determinant Polynomial}

The next polynomial of interest in the current paper is
rectangular permanent polynomial. Given a $k\times n$
rectangular matrix $\ex=(x_{i,j})_{1\leq i\leq k, 1\leq j\leq n}$ of commuting variables and a $k\times n$
rectangular matrix $\ey=(y_{i,j})_{1\leq i\leq k, 1\leq j\leq n}$ of noncommuting variables, the
rectangular permanent polynomial in commutative and noncommutative domains are defined as follows
 \[
\rper(\ex) = \sum_{\sigma \in I_{k,n}} \prod_{i=1}^k x_{i,\sigma(i)}, \quad\quad\quad\quad\quad \rper(\ey) = \sum_{\sigma \in I_{k,n}} \prod_{i=1}^k y_{i,\sigma(i)}.
\]

Here, $I_{k,n}$ is the set of all injections from $[k]\rightarrow
[n]$. An alternative view is that $\rper(\ex) = \sum_{S\subset [n] :
  |S|=k} \per(\ex_S)$ where $\ex_S$ is the $k\times k$ submatrix where the columns are indexed
by the set $S$. 
Of course, such a polynomial can be computed
in time $O^{*}(n^k)$ using a circuit of similar size, the main
interesting issue is to understand whether the dependence on the
parameter $k$ can be improved. It is implicit in the work of
Vassilevska and Williams \cite{WW13} that the $\rper(\ex)$ polynomial
in the commutative setting can be computed by an algebraic branching
program of size $O^{*}(2^k)$. This problem originates from its
connection with combinatorial problems studied in the context of exact
algorithm design \cite{WW13}. 
In the noncommutative setting, set-multilinearizing  $S_{n,k}^{*}(\ey)$ polynomial (i.e. replacing each $y_i$ at position $j$ by $y_{j,i}$), we obtain $\rper(\ey)$ where $\ey$ is a $k\times n$ symbolic matrix of noncommuting variables. Using this connection with the
explicit construction of $S_{n,k}^{*}(\ey)$ polynomial, we provide an
ABP for $\rper(\ey)$ in the \emph{noncommutative setting} of size
$O^{*}(\binom{n}{\downarrow k/2})$. The construction time is also similar.

As in the usual commutative case, the noncommutative determinant polynomial of a
symbolic matrix $\ey=(y_{i,j})_{1\leq i,j\leq k}$ is defined as
follows (the variables in the monomials are ordered from left to
right):
 \[  
 \Det(\ey)= \sum_{\sigma\in S_k} y_{1,\sigma(1)}\ldots y_{k,\sigma(k)}.
 \]
 Nisan \cite{Ni91} has also shown that any algebraic branching program
for the \emph{noncommutative determinant} of a $k\times k$ symbolic
matrix must be of size $\Omega(2^k)$. In this paper we give an
explicit construction of such an ABP in time $O^{*}(2^k)$. Here too, the main point 
is that Nisan has also shown that the lower bound is tight, but we provide an explicit 
construction. 

Moreover, motivated by the result of Vassilevska and Williams \cite{WW13}, we 
study the complexity of the rectangular determinant polynomial (in commutative domain)
defined as follows.
\[
\rdet(\ex)=\sum_{S\in {[n]\choose k}}\Det(\ex_S).
\]
The above definition is well-known in mathematics. It is often
referred to as the Cullis determinant \cite{NY07}.  
We prove that the rectangular determinant polynomial can be
computed using $O^{*}(2^k)$-size explicit ABP. Whether one can explicitly obtain 
a branching program of similar size for rectangular determinant polynomial in the noncommutative 
domain, remains as an open problem. 

Finally, we consider the problem of evaluating the noncommutative
rectangular determinant over matrix algebras and show that it is
$\#\w[1]$-hard for polynomial dimensional matrices. Hence the
noncommutative rectangular determinant is unlikely to have an explicit
$O^{*}(n^{o(k)})$-size ABP. Recently, we have shown the
$\#\w[1]$-hardness of computing rectangular permanents over
poly-dimensional rational matrices ~\cite{ACDM18}. We note that the
noncommutative $n\times n$ determinant over matrix algebras is
well-studied, and computing it remains $\# \smallp$-hard even over
$2\times 2$ rational matrices~\cite{AS10, CHSS11, Bla13}. 
Our proof technique is
based on Hadamard product of noncommutative polynomials which is also used in~\cite{AS10}.
 However, the crucial difference is that, to show
the $\#\smallp$-hardness of noncommutative determinant,
authors in~\cite{AS10} reduce the evaluation of commutative permanent to this case; whereas,
 $\#\w[1]$-the hardness of noncommutative
rectangular determinant seems more challenging as commutative rectangular permanent is in FPT.
In contrast,
we show that the rectangular determinant (and rectangular permanent),
whose entries are $r\times r$ matrices over any field, can be computed
in time $O^*(2^k r^{2k})$.

\subsection*{Our Results}
We first formally define what we mean by \emph{explicit} circuit upper
bounds.

\begin{definition}[Explicit Circuit Upper Bound]\label{explicit-def}
A family $\{f_n\}_{n>0}$ of degree-$k$ polynomials in the commutative
ring $\F[x_1,x_2,\ldots,x_n]$ (or the noncommutative ring
$\F\angle{y_1,y_2,\ldots,y_n}$) has \emph{$q(n,k)$-explicit upper
  bounds} if there is an $O^*(q(n,k))$ time-bounded algorithm
$\mathcal{A}$ that on input $\angle{0^n,k}$ outputs a circuit $C_n$ of
size $O^*(q(n,k))$ computing $f_n$.
\end{definition}

We show the following  explicit upper bound results.

\begin{theorem}\label{explicit-ub}
\begin{enumerate}
\item The family of symmetrized elementary polynomials
  $\{S^*_{n,k}(\ey)\}_{n>0}$ has ${n\choose {\downarrow
      k/2}}$-explicit ABPs over any field.
      
 \item The noncommutative rectangular permanent family
  $\{\rper(\ey)\}_{n>0}$, where $\ey$ is a $k\times n$ symbolic matrix of
  variables has ${n\choose {\downarrow k/2}}$-explicit ABPs.
  \end{enumerate}
\end{theorem}
\begin{remark}
  We note here that there is an algorithm of run time $O^*({n\choose
    {\downarrow k/2}})$ for computing the rectangular permanent over
  rings and semirings \cite{BHKK10}. Our contribution in 
  Theorem~\ref{explicit-ub}.2 is that we obtain an ${n\choose {\downarrow
      k/2}}$-explicit ABP for it.
\end{remark}


\begin{theorem}\label{abp-crdet}
\begin{enumerate}

\item The family of noncommutative determinants $\{\Det(\ey)\}_{k>0}$
  has $2^k$-explicit ABPs over any field.
  
\item There is a family $\{f_n\}$ of noncommutative degree-$k$
  polynomials $f_n$ such that $f_n$ has
  the same support as $S^*_{n,k}$, and it has $2^k$-explicit ABPs. 
  This result holds over any field that has at least $n$ distinct elements. 
 
\item  The commutative rectangular determinant family
  $\{\rdet(\ex)\}_{k>0}$, where $\ex$ is a $k\times n$ matrix of
  variables has $2^k$-explicit ABPs.

\end{enumerate}
\end{theorem} 

We stress here that the constructive aspect of the above upper bounds
is new. The \emph{existence} of the ABPs claimed in the first two
parts of Theorem~\ref{explicit-ub} and the first part of Theorem \ref{abp-crdet} follows from Nisan's work
\cite{Ni91} which shows a tight connection between optimal ABP-size
for some $f\in\F\angle{\ex}$ and ranks of the matrices $M_r$ whose rows
are labeled by degree $r$ monomials, columns by degree $k-r$ monomials
and the $(m_1,m_2)^{th}$ entry is the coefficient of $m_1m_2$ in
$f$. But constructing an ABP for $f$ would be substantially slower in
general (for example, we could adapt the Beimel et al. algorithm for
learning multiplicity automata \cite{beim} to solve this
problem). 

Next we describe the parameterized hardness result for rectangular
determinant polynomial when we evaluate over matrix algebras.

\begin{theorem}\label{ncdet-hardness-thm}
For any fixed $\epsilon>0$, evaluating the $k\times n$ rectangular
determinant polynomial over
$n^\epsilon\times n^\epsilon$ rational matrices is $\#\w[1]$-hard,
treating $k$ as fixed parameter.
\end{theorem}

However, we can easily design an algorithm of run time $O^{*}(2^k r^{2k})$
for computing the rectangular permanent and determinant polynomials
with $r\times r$ matrix entries over any field. 

\subsection*{Organization}
The paper is organized as follows. In Section \ref{prelim}, we provide
the necessary background. The proofs of Theorem \ref{explicit-ub} and
Theorem \ref{abp-crdet} are given in Section
\ref{explicit-construction} and Section \ref{sec-abp-crdet} repectively. We prove Theorem \ref{ncdet-hardness-thm}
in Section \ref{hardness}.

\section{Preliminaries} \label{prelim}

We provide some background results from noncommutative computation.
Given a commutative circuit $C$, we can naturally associate a
noncommutative circuit $C^{nc}$ by prescribing an input order at each
multiplication gate. This is captured in the following definition.

\begin{definition}\label{noncomm-version}
Given a commutative circuit $C$ computing a polynomial in $\F[x_1,
  x_2, . . . , x_n]$, the noncommutative version of $C$, $C^{nc}$ is
the noncommutative circuit obtained from $C$ by fixing an ordering of
the inputs to each product gate in $C$ and replacing $x_i$ by the
noncommuting variable $y_i : 1\leq i \leq n$.
\end{definition}

Let $f \in \F[\ex]$ be a homogenous degree-$k$ polynomial computed by a circuit 
$C$, and let $\hat{f}(\ey)\in\F\angle{\ey}$ be the polynomial computed by $C^{nc}$.  
Let $\ex_k$ denote the set of all degree-$k$ monomials over $\ex$. As
usual, $\ey^k$ denotes all degree-$k$ noncommutative monomials (i.e.,
words) over $\ey$. Each monomial $m\in \ex_k$ can appear as different
noncommutative monomials $\hat{m}$ in $\hat{f}$. We use the notation
$\hat{m}\to m$ to denote that $\hat{m}\in \ey^k$ will be transformed to
$m\in X_k$ by substituting $x_i$ for $y_i, 1\le i\le n$. Then, we
observe the following, $[m]f = \sum_{\hat{m}\to m}[\hat{m}]\hat{f}.$

For each monomial $\hat{m}=y_{i_1}y_{i_2}\cdots y_{i_k}$, the
permutation $\sigma\in S_k$ maps $\hat{m}$ to the monomial
$\hat{m}^{\sigma}$ defined as
$\hat{m}^{\sigma}=y_{i_{\sigma(1)}}y_{i_{\sigma(2)}}\cdots
y_{i_{\sigma(k)}}$. By linearity, $\hat{f}=\sum_{\hat{m}\in
  \ey^k}[\hat{m}]\hat{f} \cdot \hat{m}$ is mapped by $\sigma$ to the
polynomial, $\hat{f}^\sigma=\sum_{\hat{m}\in \ey^k}[\hat{m}]\hat{f}\cdot
\hat{m}^\sigma$. This gives the following definition. 

\begin{definition}\label{symmetric}
The \emph{symmetrized polynomial} of $f$, $f^*$ is degree-$k$ homogeneous
polynomial $f^* = \sum_{\sigma\in S_k} \hat{f}^{\sigma}$.
\end{definition}

Next, we recall the definition of Hadamard product of two polynomials. 

\begin{definition}\label{had-product}
  Given polynomials $f, g$, their Hadamard product is defined as
\[
f\circ g = \sum_{m} ([m]f \cdot [m]g) \cdot m,
\] 
where $[m]f$ denotes the coefficient of monomial $m$ in $f$.
\end{definition}

In the commutative setting, computing the Hadamard product is
intractable in general. This is readily seen as the Hadamard product
of the determinant polynomial with itself yields the permanent
polynomial. However, in the noncommutative setting the Hadamard
product of two ABPs can be computed efficiently \cite{AJS09}.

\begin{theorem}\label{abp-abp}{\rm\cite{AJS09}} 
  Given a noncommutative ABP of size $S'$ for degree $k$ polynomial $f
  \in \F\angle{y_1, y_2, \ldots, y_n}$ and a noncommutative ABP of
  size $S$ for another degree $k$ polynomial $g \in \F\angle{y_1, y_2,
    . . . , y_n}$, we can compute a noncommutative ABP of size $SS'$
  for $f \circ g$ in deterministic $SS'\cdot \poly(n, k)$ time.
\end{theorem}

Let $C$ be a circuit and $B$ an ABP computing homogeneous degree-$k$
polynomials $f,g\in\F\langle{\ey \rangle}$ respectively.  Then their
Hadamard product $f\circ g$ has a noncommutative circuit of
polynomially bounded size which can be computed efficiently
\cite{AJS09}.

Furthermore, if $C$ is given by black-box access then $f\circ
g(a_1,a_2,\ldots,a_n)$ for $a_i\in\F, 1\le i\le n$ can be evaluated by
evaluating $C$ on matrices defined by the ABP $B$ \cite{AS10} as
follows: For each $i \in [n]$, the transition matrix $M_i \in
M_{s}(\F)$ are computed from the noncommutative ABP $B$ (which is of
size $s$) that encode layers. We define $ M_i[k,\ell] = [x_i]
L_{k,\ell}, $ where $L_{k,\ell}$ is the linear form on the edge
$(k,\ell)$.  Now to compute $(f\circ g) (a_1,a_2,\ldots,a_n)$ where
$a_i \in \F$ for each $1\leq i \leq n$, we compute $C(a_1 M_1,a_2
M_2,\ldots a_n M_n)$.  The value $(f \circ g) (a_1,a_2,\ldots,a_n)$ is
the $(1,s)^{th}$ entry of the matrix $f(a_1 M_1,a_2 M_2,\ldots, a_n
M_n)$. 

\begin{lemma}{\rm\cite{AS10}}\label{matrix-valued}
Given a circuit $C$ and a ABP $B$ computing homogeneous noncommutative
polynomials $f$ and $g$ in $\F\langle\ey\rangle$, the Hadamard product
$f\circ g$ can be evaluated at any point $(a_1, \ldots, a_n)\in\F^n$
by evaluating $C(a_1 M_1, \ldots, a_n M_n)$ where $M_1, \ldots, M_n$
are the transition matrices of $B$, and the dimension of each $M_i$ is
the size of $B$.
\end{lemma}  

\section{The Proof of Theorem~\ref{explicit-ub}}\label{explicit-construction}

In this section, we present the construction of explicit ABPs
for $S^*_{n,k}(\ey)$ and noncommutative $\rper(\ey)$.

\subsection{The construction of ABP for $S^{*}_{n,k}(\ey)$}\label{explicit-symm-symmetric}

The construction of the ABP for $S^*_{n,k}(\ey)$ is inspired by a
inclusion-exclusion based dynamic programming algorithm for the
disjoint sum problem \cite{BHKK09}. The main result of this section is
the following.

\begin{tjoneproof}
  Let us denote by $F$ the family of subsets of $[n]$ of size exactly
  $k/2$. Let $\daF$ denote the family of subsets of $[n]$ of size at
  most $k/2$.  For a subset $S \subset [n]$, we define $m_S =
  \prod_{j\in S} y_j$.  Let us define
 \[
 f_S=\sum_{\sigma \in S_{k/2}} \prod^{k/2}_{j=1} y_{i_{\sigma(j)}}
 \]
 where $S\in F$ and $S=\{i_1 , i_2,\ldots,i_{k/2} \}$, otherwise for
 subsets $S\notin F$, we define $f_S = 0$.  Note that, for each $S\in
 F$, $f_S$ is the symmetrization of the monomial $m_S$ which we denote by $m_S^{*}$ 
 (notice Definition \ref{symmetric}).
 
 For each $S\in \daF$, let us define $\hat{f}_S = \sum_{S \subseteq A}
 f_A$ where $A\in F$. We now show, using the inclusion-exclusion
 principle, that we can express $S^*_{n,k}$ using an appropriate
 combination of these symmetrized polynomials for different subsets.
 
 \begin{lemma}\label{incexc}
  \[S^*_{n,k}=\sum_{S\in \daF} (-1)^{|S|} {\hat{f}_S}^2.\]
 \end{lemma}
 \begin{proof}
 Let us first note that, $S^*_{n,k} = \sum_{A\in F} \sum_{B\in F}
 [A\cap B = \emptyset] f_A f_B$, where we use $[P]$ to
 denote that the proposition $P$ is true. By the inclusion-exclusion
 principle:
 \begin{align*}
    S^*_{n,k} &= \sum_{A\in F} \sum_{B\in F} [A\cap B = \emptyset] f_A f_B\\
    &=\sum_{A\in F} \sum_{B\in F} \sum_{S\in \daF} (-1)^{|S|}[S\subseteq A\cap B] f_A f_B\\
    &=\sum_{S\in \daF} (-1)^{|S|}\sum_{A\in F} \sum_{B\in F} [S\subseteq A][S\subseteq B] f_A f_B\\
    &=\sum_{S\in \daF} (-1)^{|S|} \left( \sum_{A\in F}  [S\subseteq A] f_A \right)^2 = \sum_{S\in \daF} (-1)^{|S|} {\hat{f}_S}^2.
\end{align*}
\end{proof}

Now we describe two ABPs where the first ABP simultaneously computes
$f_A$ for each $A\in F$ and the second one simultaneously computes
$\hat{f}_S$ for each $S\in \daF$.

\begin{lemma}\label{part0}
  There is an $\binom{n}{\downarrow {k/2}}$-explicit multi-output
  ABP $B_1$ that outputs the collection $\{ f_A \}$ for each $A \in
  F$.
\end{lemma}
\begin{proof}
  First note that, $m^*_S = \sum_{j\in S} m^*_{S \setminus \{j\}}\cdot
  x_j$. Now, the construction of the ABP is obvious. It consists of
  $(k/2+1)$ layers where layer $\ell\in \{0,1,\ldots,k/2\}$ has
  $\binom{n}{\ell}$ many nodes indexed by $\ell$ size subsets of
  $[n]$. In $(\ell+1)^{th}$ layer, the node indexed by $S$ is connected
  to the nodes $S\setminus \{j\}$ in the previous layer with an edge
  label $x_j$ for each $j\in S$. Clearly, in the last layer, the
  $S^{th}$ sink node computes $f_S$.
\end{proof}

\begin{lemma}\label{part1}
  There is an $\binom{n}{\downarrow {k/2}}$-explicit multi-output
  ABP $B_2$ that outputs the collection $\{ \hat{f}_S \}$ for each $S
  \in \daF$.
\end{lemma}

\begin{proof}
  To construct such an ABP, we use ideas from \cite{BHKK09}. We define
  $\hat{f}_{i,S} = \sum_{S \subseteq A} f_A$ where $S\subseteq A$ and
  $A\cap [i] = S\cap [i]$. Note that, $\hat{f}_{n,S} = f_S$ and
  $\hat{f}_{0,S} = \hat{f}_S$. From the definition, it is clear that
  $\hat{f}_{i-1,S} =\hat{f}_{i,S} + \hat{f}_{i,S\cup \{i\}}$ if
  $i\notin S$ and $\hat{f}_{i-1,S} =\hat{f}_{i,S}$ if $i\in S$. Hence,
  we can take a copy of ABP $B_1$ from Lemma~\ref{part0}, and then
  simultaneously compute $\hat{f}_{i,S}$ for each $S\in \daF$ and $i$
  ranging from $n$ to $0$. Clearly, the new ABP $B_2$ consists of
  $(n+k/2+1)$ many layers and at most $\binom{n}{\downarrow{k/2}}$
  nodes at each layer. The number of edges in the ABP is also linear
  in the number of nodes.
\end{proof}

Let $f = \sum_{m\in Y^k} [m]f\cdot m$ be a noncommutative polynomial
of degree $k$ in $\F\angle{Y}$. The \emph{reverse} of $f$ is defined
as the polynomial
\[
f^R = \sum_{m\in Y^k} [m]f\cdot m^R,
\] 
where $m^R$ is the reverse of the word $m$.

\begin{lemma}\label{rev}[Reversing an ABP]
  Suppose $B$ is a multi-output ABP with $r$ sink nodes where the
  $i^{th}$ sink node computes $f_i \in \F\angle{\ey}$ for each $i\in
  [r]$. We can construct an ABP of twice the size of $B$ that computes
  the polynomial $\sum^{r}_{i=1} f_i \cdot L_i \cdot f^R_i$ where
  $L_i$ are affine linear forms.
\end{lemma}
\begin{proof}
  Suppose $B$ has $\ell$ layers, then we construct an ABP of $2\ell +1$
  layers where the first $\ell$ layers are the copy of ABP $B$ and the
  last $\ell$ layers are the ``mirror image" of the ABP $B$, call it
  $B^R$. In the $(\ell + 1)^{th}$ layer we connect the $i$th sink node of
  ABP $B$ to the $i$th source node of $B^R$ by an edge with edge label
  $L_i$. Note that, $B^R$ has $r$ source nodes and one sink node and
  the polynomial computed between $i$th source node and sink is
  $f^R_i$.
\end{proof}

Now, applying the construction of Lemma~\ref{rev} to the multi-output
ABP $B_2$ of Lemma~\ref{part1} with $L_S = (-1)^{|S|}$ we obtain an
ABP that computes the polynomial $\sum_{S} (-1)^{|S|} \hat{f}_S \cdot
\hat{f}^R_S$.  Since $\hat{f}_S$ is a symmetrized polynomial, we note
that $\hat{f}^R_S = \hat{f}_S$ and using Lemma~\ref{incexc} we
conclude that this ABP computes $S^*_{n,k}$.  The ABP size is
$O(k\binom{n}{\downarrow{k/2}})$. 
\qed
\end{tjoneproof} 
 
\subsection{The construction of ABP for $\rper(\ey)$} 
 
 \begin{tjtwoproof}
A $\binom{n}{\downarrow {k/2}}$-explicit ABP for the rectangular
permanent polynomial can be obtained easily from the
$\binom{n}{\downarrow {k/2}}$-explicit ABP for $S^{*}_{n,k}(\ey)$ by
careful set-multilinearization. This can be done by simply renaming
the variables $y_i : 1\leq i\leq n$ at the position $1\leq j\leq k$ by
$y_{j,i}$. 
\qed
\end{tjtwoproof} 
 
 \section{The Proof of Theorem~\ref{abp-crdet}}\label{sec-abp-crdet}
We divide the proof in three subsections. 

\subsection{A $2^k$-explicit ABP for $k\times k$ noncommutative
  determinant}\label{ncd-explicit}

In this section, we present an optimal explicit ABP construction for the
noncommutative determinant polynomial for the square symbolic matrix. .

\begin{tkoneproof}
  The ABP $B$ has $k+1$ layers with ${k \choose \ell}$ nodes at the
  layer $\ell$ for each $0\leq \ell\leq k$. The source of the $\ABP$
  is labeled $\emptyset$ and the nodes in layer $\ell$ are labeled by
  the distinct size $\ell$ subsets $S\subseteq [k]$, $1\le \ell \le
  k$, hence the sink is labeled $[k]$. From the node labeled $S$ in
  layer $\ell$, there are $k-\ell$ outgoing edges $(S,S\cup\{j\})$,
  $j\in [k]\setminus S$.

  Define the sign $\sgn(S,j)$ as $ \sgn(S,j)=(-1)^{t_j}$, where $t_j$
  is the number of elements in $S$ larger than $j$.  Equivalently,
  $t_j$ is the number of swaps required to insert $j$ in the correct
  position, treating $S$ as a sorted list.

  For noncommutative determinant polynomial, we connect the set $S$ in
  the $i^{th}$ layer to a set $S\cup \{j\}$ in the $(i+1)^{th}$ layer with
  the edge label $\sgn(S,j) \cdot x_{i+1, j}$ The source to sink paths
  in this ABP are in 1-1 correspondence to the node labels on the
  paths which give subset chains $\emptyset\subset T_1\subset
  T_2\subset\cdots \subset T_k=[k]$ such that $|T_i\setminus
  T_{i-1}|=1$ for all $i\le k$. Such subset chains are clearly in 1-1
  correspondence with permutations $\sigma\in S_k$ listed as a
  sequence: $\sigma(1),\sigma(2),\ldots,\sigma(k)$, where
  $T_i=\{\sigma(1),\sigma(2),\ldots,\sigma(i)\}$. The following claim
  spells out the connection between the sign $\sgn(\sigma)$ of
  $\sigma$ and the $\sgn(S,j)$ function defined above.

\begin{claim}\label{sign}
  For each $\sigma\in S_k$ and
  $T_i=\{\sigma(1),\sigma(2),\ldots,\sigma(i)\}$, we have
\[
\sgn(\sigma) = \prod_{i=1}^k \sgn(T_{i-1},\sigma(i)).
\]
\end{claim}

\claimproof{We first note that $\sgn(\sigma) = (-1)^t$, if there are
  $t$ transpositions $(r_i~s_i), 1\le i\le t$ such that $\sigma\cdot
  (r_1~s_1)\cdot (r_2~s_2)\cdots (r_t~s_t)=1$.  Equivalently,
  interpreting this as sorting the list
  $\sigma(1),\sigma(2),\ldots,\sigma(k)$ by swaps $(r_i~s_i)$,
  applying these $t$ swaps will sort the list into $1,2,\ldots,k$. As
  already noted, $\sgn(T_{i-1},\sigma(i))=(-1)^{t_i}$, where $t_i$ is
  the number of swaps required to insert $\sigma(i)$ in the correct
  position into the sorted order of $T_{i-1}$ (where $\sigma(i)$ is
  initially placed to the right of $T_{i-1}$). Hence, $\sum_{i=1}^k
  t_i$ is the total number of swaps required for this insertion sort
  procedure to sort $\sigma(1),\sigma(2),\ldots,\sigma(k)$. It follows
  that $\prod_{i=1}^k \sgn(T_{i-1},\sigma(i))=(-1)^{\sum_{i}
    t_i}=\sgn(\sigma)$, which proves the claim.}
\qed

The fact that the ABP computes the noncommutative determinant
polynomial follows directly from Claim \ref{sign} and the edge labels.
\qed
\end{tkoneproof}

\subsection{A $2^k$-explicit ABP weakly equivalent to
  $S^*_{n,k}$}\label{explicit-weakly-equivalent}

A polynomial $f\in \F[\ex]$ (resp. $\F\angle{\ey}$) is said to be
\emph{weakly equivalent} to a polynomial $g\in \F[\ex]$
(resp.~$\F\angle{\ey}$), if for each monomial $m$ over $X$, $[m]f = 0$
if and only if $[m]g = 0$. For the construction of an ABP computing a
polynomial weakly equivalent to $S_{n,k}^{*}$, we will suitably modify
the ABP construction described above.

\begin{tktwoproof}
Let $\alpha_i, 1\le i\le n$ be distinct elements from $\F$. For each
$j\in [k]\setminus S$, the edge $(S,S\cup\{j\})$ is labeled by the
linear form $\sgn(S,j)\cdot \sum^n_{i=1} \alpha^j_i x_i$, where $x_i,
1\le i\le n$ are noncommuting variables.  This gives an $\ABP$ $B$ of
size $O^*(2^k)$.

We show that the polynomial computed by ABP $B$ is weakly equivalent
to $S^*_{n,k}$.  Clearly, $B$ computes a homogeneous degree $k$
polynomial in the variables $x_i, 1\le i\le n$. We determine the
coefficient of a monomial $x_{i_1}x_{i_2}\cdots x_{i_k}$. As noted,
each source to sink path in $B$ corresponds to a permutation
$\sigma\in S_k$.  Along that path the ABP compute the product of
linear forms
\[
\sgn(\sigma) L_{\sigma(1)}L_{\sigma(2)}\cdots L_{\sigma(k)}, \textrm{ where } 
L_{\sigma(q)} = \sum_{i=1}^n\alpha_i^{\sigma(q)} x_i,
\]
where the sign is given by the previous claim. The coefficient of
monomial $x_{i_1}x_{i_2}\cdots x_{i_k}$ in the above product is given
by $\sgn(\sigma)\prod_{q=1}^k\alpha_{i_q}^{\sigma(q)}$.  Thus, the
coefficient of $x_{i_1}x_{i_2}\cdots x_{i_k}$ in the ABP is given by
$\sum_{\sigma\in
  S_k}\sgn(\sigma)\prod_{q=1}^k\alpha_{i_q}^{\sigma(q)}$, which is the
determinant of the $k\times k$ Vandermonde matrix whose $q^{th}$
column is
$(\alpha_{i_q},\alpha_{i_q}^2,\ldots,\alpha_{i_q}^{k})^T$. Clearly,
that determinant is non-zero if and only if the monomial
$x_{i_1}x_{i_2}\cdots x_{i_k}$ is multilinear. Clearly the proof works 
for any field that contains at least $n$ distinct elements. 
\qed
\end{tktwoproof}

\begin{remark}\label{main-open}
  A polynomial $f\in \F\angle{\ey}$ is \emph{positively weakly
    equivalent} to $S^*_{n,k}$, if for each multilinear monomial $m\in
  \ey^k$, $[m]f >0$. In the above proof,  let $g$ be the
  polynomial computed by ABP $B$ that is weakly equivalent to
  $S^*_{n,k}$. Clearly, $f=g\circ g$ is positively weakly equivalent
  to $S^*_{n,k}$, and $f$ has a $4^k$-explicit ABP, since $B$ is
  $2^k$-explicit. This follows from Theorem \ref{abp-abp}. We leave open the problem of finding a
  $2^k$-explicit ABP for some polynomial that is positively weakly
  equivalent to $S^*_{n,k}$. Such an explicit construction would imply a
  deterministic $O^{*}(2^k)$ time algorithm for $k$-path which is a
  long-standing open problem \cite{KW16}.
\end{remark}

\subsection{A $2^k$-explicit ABP for $k\times n$ commutative
  rectangular determinant}\label{rdet-explicit-abp}

In this section, we present the ABP construction for commutative determinant 
polynomial for $k\times n$ symbolic matrix. 

\begin{tkthreeproof}
  We adapt the ABP presented in Subsection \ref{ncd-explicit}. The
  main difference is that, for the edge $(S,S\cup\{j\})$, the linear
  form is
  $\sgn(S,j)\cdot\left(\sum_{i=1}^n x_{j, i} z_i\right)$, where $z_i :
  1\leq i\leq n$ are fresh noncommuting variables, and the $x_{j,i} :
  1\leq j\leq k, 1\leq i\leq n$ are commuting variables.

Then with a similar argument as before, the coefficient of the
monomial $z_{i_1} z_{i_2} \ldots z_{i_k}$ where $i_1<i_2<\ldots<i_k$
is given by $\sum_{\sigma\in S_k} \sgn(\sigma) x_{\sigma(1), i_1}
\ldots x_{\sigma(k), i_k}$ Now for a fixed $\sigma\in S_k$, let
$\tau^{\sigma}$ be the injection $[k]\rightarrow [n]$ such that
$\tau^{\sigma}(j)=i_{\sigma^{-1}(j)} : 1\leq j\leq k$.

Let $(j_1, j_2)$ be an index pair that is an inversion in $\sigma$,
i.e.  $j_1 < j_2$ and $\sigma(j_1) > \sigma(j_2)$. Let $\ell_1 =
\sigma(j_1)$ and $\ell_2 = \sigma(j_2)$. So $i_{\tau^{\sigma}(\ell_1
  )} = i_{\sigma^{-1}(\ell_1)}$ and $i_{\tau^{\sigma}(\ell_2 )} =
i_{\sigma^{-1}(\ell_2)}$. Clearly, $i_{\tau^{\sigma}(\ell_1 )} <
i_{\tau^{\sigma}(\ell_2 )}$. Hence:

\[
\sum_{\sigma\in S_k} \sgn(\sigma) x_{\sigma(1), i_1} \ldots x_{\sigma(k), i_k}=\sum_{\tau^{\sigma}\in I_{k,n}} \sgn(\tau^{\sigma}) x_{1, \tau^{\sigma}(1)} \ldots x_{k, \tau^{\sigma}(k)}.  
\]
 
Now the idea is to filter out only the good monomials $z_{i_1} z_{i_2}
\ldots z_{i_k}$ where $i_1 < i_2 <\ldots <i_k$ from among all the
monomials.  This can be done by taking Hadamard product (using Theorem
\ref{abp-abp}) with the following polynomial,
\[
S^{nc}_{n,k}(\ez) = \sum_{S=\{i_1 < i_2 <\ldots < i_k\}}  z_{i_1} z_{i_2} \ldots z_{i_k}. 
\] 
Clearly, $S^{nc}_{n,k}$ has a $\poly(n,k)$-sized ABP which is just the
noncommutative version (see Definition \ref{noncomm-version}) of the
well-known ABP for commutative $S_{n,k}$.
Finally, we substitute each $z_i=1$ to get the desired ABP for
$\rdet(\ex)$.
\qed
\end{tkthreeproof}

\section{Hardness of Evaluating Rectangular Determinant Over Matrix Alegbras}\label{hardness}

In this section we prove a hardness result for evaluating the
rectangular determinant over matrix algebras. More precisely, if $A$
is a $k\times n$ matrix whose entries $A_{ij}$ are $n^{\epsilon}\times
n^{\epsilon}$ rational matrices for a fixed $\epsilon>0$, then it is
$\#\w[1]$-hard to compute $\rdet(A)$. We show this by a reduction from
the $\#\w[1]$-complete problem of counting the number of simple
$k$-paths in directed graphs.

However, there is a simple algorithm of run time $O^{*}(2^k r^{2k})$ to
evaluate rectangular permanent or rectangular determinant of size
$k\times n$ over matrix algebras of dimension $r$.  The proof is given
in the appendix.

For the proof of Theorem~\ref{ncdet-hardness-thm}, we also use the notion of \emph{Graph Polynomial}.  Let $G(V,E)$ be a
directed graph with $n$ vertices where $V(G)= \{v_1,v_2,\ldots,
v_n\}$.  A $k$-walk is a sequence of $k$ vertices $v_{i_1},
v_{i_2},\ldots, v_{i_k}$ where $(v_{i_j},v_{i_{j+1}})\in E$ for each
$1\leq j \leq k-1$. A $k$-path is a $k$-walk where no vertex is
repeated.  Let $A$ be the adjacency matrix of $G$, and let
$z_1,z_2,\ldots,z_n$ be noncommuting variables. Define an $n\times n$
matrix $B$
\[
B[i, j] = A[i, j]\cdot z_i,~~ 1\leq i,j \leq n.
\]
Let $\vec{1}$ denote the all $1$'s vector of length $n$. Let $\vec{z}$
be the length $n$ vector defined by $\vec{z}[i] = z_i$. The
\emph{graph polynomial} $C_G\in\F\angle{\ez}$ is defined as
\[
C_G(z_1,z_2,\ldots,z_n) = \vec{1}^T\cdot B^{k-1}\cdot \vec{z}.
\]
Let $W$ be the set of all $k$-walks in $G$. The following observation
is folklore.

\begin{obs} \label{obs-graph-poly}
\[
C_G(z_1,z_2,\ldots,z_n) =  \sum_{v_{i_1}v_{i_2}\ldots v_{i_k}\in W} z_{i_1}z_{i_2}\cdots z_{i_k}.	
\]
Hence, $G$ contains a $k$-path if and only if the graph polynomial
$C_G$ contains a multilinear term.
\end{obs}

\subsection{The Proof of Theorem \ref{ncdet-hardness-thm}}
Let $I_{k,n}$ be the set of injections from $[k]\rightarrow [n]$.
Define
 \begin{align*}
 S := \{f \in I_{2k,2n} | \exists g\in I_{k,n} \text{ such that}~ \forall i\in[k], f(2i-1) = g(i) ; f(2i) = n + g(i) \}. 
 \end{align*}
 
Clearly, there is a bijection between $S$ and $I_{k,n}$. We denote
each $f\in S$ as $f_g$ where $g\in I_{k,n}$ is the corresponding
injection. By a simple counting argument, we observe the
following.
 
\begin{obs}\label{obs-easy}
  For each $f\in S, \sgn(f) = (-1)^{\frac{k(k-1)}{2}}$.
 \end{obs}

Consider a set of noncommuting variables
$\ey=\{y_{1,1},y_{1,2},\ldots,y_{2k,2n}\}$ corresponding to the
entries of a $2k\times 2n$ symbolic matrix $\ey$.  Given $f\in
I_{2k,2n}$, define $m_f = \prod_{i=1}^{2k} y_{i,f(i)}$.

\begin{lemma}\label{filter}
 There is an ABP $B$ of $\poly(n,k)$ size that computes a polynomial
 $F\in \F\angle{\ey}$ such that for each $f\in I_{2k,2n}$, $[m_f]F = 1$
 if $f\in S$ and otherwise $[m_f]F = 0$.
\end{lemma}

\begin{proof}
 The ABP $B$ consists of $2k+1$ layers, labelled
 $\{0,1,\ldots,2k\}$. For each even $i\in [0,2k]$, there is exactly
 one node $q_i$ at level $i$. For each odd $i\in[0,2k]$, there are $n$
 nodes $p_{i,1},p_{i,2},\ldots,p_{i,n}$ at level $i$. We now describe
 the edges of $B$. For each even $i\in [0,2k-2]$ and $j\in [n]$, there
 is an edge from $q_i$ to $p_{i+1,j}$ labelled $x_{i+1,j}$. For each
 odd $i\in [0,2k-1]$ and $j\in[n]$, there is an edge from $p_{i,j}$ to
 $q_{i+1}$ labelled $x_{i+1,n+j}$. For an injection $f\in I_{2k,2n}$,
 $B$ contributes a monomial $m_{f}$ if and only if $f\in S$ and $B$
 can be computed in $\poly(n,k)$ time.
\end{proof}

Suppose, $\ey$ is a $2k\times 2n$ matrix where the $(i,j)^{th}$ entry is
$y_{i,j}$. By Observation \ref{obs-easy} and Lemma \ref{filter},
\[\rdet(\ey)\circ F(\ey) = \sum_{f_g\in S} {\sgn(f_g)} m_{f_g} = (-1)^{\frac{k(k-1)}{2}} \sum_{g\in I_{k,n}} m_{f_g}.
\]

Let $Z = \{z_1,\ldots,z_n\}$ be a set of noncommuting variables. Define for each $g\in I_{k,n}$, $m'_g = \prod_{i=1}^k z_{g(i)}$.
Define a map $\tau$ such that $\tau: y_{i,j}\mapsto z_{j}$ if $i$ is
odd, and $\tau: y_{i,j}\mapsto 1$ for even $i$. 
In other words, $\tau(m_{f_g}) = m'_g$.
Notice that,
\[\rdet(\ey)\circ F(\ey)|_{\tau} = {(-1)^{\frac{k(k-1)}{2}}} \sum_{g\in I_{k,n}} {m_{f_g}|_{\tau}} = {(-1)^{\frac{k(k-1)}{2}}} \sum_{g\in I_{k,n}} {m'_g}={(-1)^{\frac{k(k-1)}{2}}} S^*_{n,k}(\ez).
\]

Given a directed graph $G$ on $n$ vertices, we first construct an ABP
for the noncommutative graph polynomial $C_G$ over rationals. From the
definition, it follows that $C_G$ has a polynomial size ABP. Notice
that,
$
((\rdet(\ey)\circ F(\ey)|_{\tau})\circ C_G(\ez))(\vec{1}) = S^*_{n,k}(\ez)\circ C_G(\ez)(\vec{1})
$ 
counts the number of directed $k$-paths in the graph $G$, and hence
evaluating this term is $\#\w[1]$-hard. Let us modify the ABP for
graph polynomial $C_G(\ez)$ by replacing each edge labeled by $z_j$ at
$i^{th}$ layer by two edges where the first edge is labeled by
$y_{2i-1,j}$ and second one is labeled by $y_{2i,n+j}$. 
Let $C'_{G}(\ey)$ is the new polynomial computed by the ABP.
Notice that, each monomial of the modified graph polynomial looks like
$\prod_{i=1}^{2k} y_{i,f(i)}$ for some $f: [2k] \mapsto [2n]$. More importantly,
for each $k$-path $v_{i_1}v_{i_2}\ldots v_{i_k}$, if $g\in I_{k,n}$ is the corresponding injection,
then $\prod_{i=1}^k z_{g(i)}$ is converted to $\prod_{i=1}^{2k} y_{i,f_g(i)}$ for $f_g\in S$.
Notice that,
$(\rdet(\ey)\circ F(\ey)|_{\tau})\circ C_G(\ez)$ $ = (\rdet(\ey)\circ
F(\ey)\circ C'_G(\ey))|_{\tau}$ and hence, evaluating $(\rdet(\ey)\circ
F(\ey)\circ C'_G(\ey))(\vec{1})$ is $\#\w[1]$-hard.

Now, assume to the contrary, we have an FPT algorithm $\mathcal{A}$ to
evaluate $\rdet(\ey)$ over matrix inputs.  As, $C'_G(\ey)$ and $F(\ey)$ are
computed by ABPs, we obtain an ABP $B'$ computing $C'_G\circ
F(\ey)$. From ABP $B'$, we construct the $t \times t$ transition
matrices $M_{1,1},\ldots,M_{2k,2n}$ where $t$ is the size of the ABP
$B'$.  From Lemma \ref{matrix-valued} we know that, we are interested
to compute $\rdet(\ey)$ over the matrix tuple $(M_{1,1},\ldots, M_{2k,2n})$ which is same as
invoking the algorithm $\mathcal{A}$ on the following $2k \times 2n$
matrix $A$: $a_{i,j} = M_{i,j}$.
By a simple reduction we get a similar hardness over $n^{\epsilon}\times n^{\epsilon}$ dimensional 
matrix algebras for any fixed $\epsilon >0$. 
\qed

%

\bibliography{ref2}

\newpage
\appendix

\section{Computing Rectangular Permanent and Determinant over Small Dimensional Algebras}\label{fpt-smalld}

The main result of the section is as follows. 

\begin{theorem}\label{rperm}
  Let $\mathbb{F}$ be any field and $\mathcal{A}$ be an $r$ dimensional
  algebra over $\mathbb{F}$ with basis $e_1 ,e_2 ,\ldots,e_r$.  Let
  $\{A_{ij} \}_{\substack{1\leq i \leq k \\ 1\leq j \leq n}}$ be a $k
  \times n$ matrix with $A_{ij} \in \mathcal{A}$.  Then $\rper(A)$ and
  $\rdet(A)$ can be computed in deterministic $O^*(2^k r^{2k})$ time.
\end{theorem}

\begin{proof}
  We present the proof for rectangular permanent. The proof for
  rectangular determinant is identical.
  The proof follows easily from expressing each entry $A_{i,j}$ in the
  standard basis and then rearranging terms.  
  Let $e_1, e_2, \ldots, e_r$ be the standard basis for $\mathcal{A}$ over $\F$. 
  First we note that

  \begin{equation}
 \begin{split}
   \rper(A) &= \sum_{\substack{f \in I_{k,n}}} \prod^{k}_{i=1} A_{i f(i)} \\
   &= \sum_{\substack{f\in I_{k,n}}} \prod^{k}_{i=1} \sum^{r}_{\ell =1 } A^{(\ell)}_{i f(i)} e_{\ell} \\
   &= \sum_{\substack{f\in I_{k,n}}}  \sum_{(t_1,t_2,\ldots,t_k) \in [r]^k} \prod^{k}_{i=1} A^{(t_i)}_{i f(i)}  \prod^{k}_{i=1} e_{t_i} \\
   &= \sum_{(t_1,t_2,\ldots,t_k) \in [r]^k} (\sum_{\substack{f \in
       I_{k,n}}} \prod^{k}_{i=1} A^{(t_i)}_{i f(i)} ) \prod^{k}_{i=1}
   e_{t_i}.
 \end{split}
 \end{equation}

 Now we observe that
 $$\sum_{\substack{f \in I_{k,n}}} \prod^{k}_{i=1} A^{(t_i)}_{i f(i)}  = \rper(A^{(t_1,t_2,\ldots,t_k)}),$$ 
 where $A^{ (t_1,t_2,\ldots,t_k) }$ is the $k \times n$ matrix defined as $A^{(t_1,t_2,\ldots,t_k)}_{ij} = A^{t_i}_{ij}$.
 Thus we have 
 \begin{align}\label{theeq}
   \rper(A) = \sum_{(t_1,t_2,\ldots,t_k) \in [r]^k}
   \rper(A^{(t_1,t_2,\ldots,t_k)}) \prod^{k}_{i=1} e_{t_i}.
 \end{align}
 
 For a fixed $(t_1,t_2,\ldots,t_k) \in [r]^k$ the value
 $\rper(A^{(t_1,t_2,\ldots,t_k)})$ can be computed in $O^*(2^k)$ time
 using the rectangular permanent algorithm \cite{WW13}.
 Now we can compute $\rper(A)$ by computing $r^k$ many such
 rectangular permanents and putting them together according to
 equation~\ref{theeq}. This gives a deterministic $O^*(2^k r^{2k})$ time
 algorithm for computing $\rper(A)$.
\end{proof}

As a direct corollary we get the following. 

\begin{corollary}
  Let $\mathbb{F}$ be any field and let $A$ be a $k \times n$ matrix
  with $A_{ij} \in \mathbb{M}_{r \times r}(\mathbb{F})$.  Then
  $\rper(A)$ and $\rdet(A)$ can be computed in $O^*(2^k r^{2k})$ time.
\end{corollary}

\end{document}